# Parallel implementation of a vehicle rail dynamical model for multi-core systems

Anas M. Al-Oraiqat [*]
[*] Taibah University, Department of Computer Sciences and Information
Kingdom of Saudi Arabia, P.O. Box 2898

*Abstract* — *This research presents a model of a complex dynamic object running on a multi-core system. Discretization and numerical integration for multibody models of vehicle rail elements in the vertical longitudinal plane fluctuations is considered. The implemented model and solution of the motion differential equations allow estimating the basic processes occurring in the system with various external influences. Hence the developed programming model can be used for performing analysis and comparing new vehicle designs.*

**Keywords-dynamic model; multi-core system; SMP system; rolling stock.**

I. INTRODUCTION

The development of multi-core processors provides to developers new possibilities for expansion and increase the complexity of computer models of dynamic systems. Dynamic modelling is a powerful tool for developing objects, supporting the early design stage, real-time simulations and monitoring [1,2].

Vehicle dynamics is one the most important factors of safety and efficiency of railways [3]. Modern simulation tools have the capability of accurately analysing the diverse aspects of vehicle behaviour at minimum cost. Manchester tests of vehicle dynamics simulation are widely recognized as standard models of railway [4]. The Manchester Benchmark for the "multibody" solution has tests-templates.

- The first test case (A-1): Though it is considered as an academic case, it is still useful for a comparison of the processes involved in how the various mathematical vehicle models are constructed.
- The second test case (A-2): It concerns a typical contact wheel to make the scenario more realistic that appears on the flanges. It demonstrates modes of vibration experimenting with the effect of damping, spring stiffness and modelling of switches and crossings.
- The third test case (A-3) is used for a typical contact at the transition between tread and flange for a realistic 2-bogie local train vehicle model.

These test cases are used for the comparison of software systems for the study of dynamics of railway vehicles [5]. Such studies are carried out by programs, considering a general multibody rail vehicle dynamics computer model, such as Automatic Dynamic Analysis of Mechanical Systems (ADAMS) Rail MSC Software [6], MEDYNA [7], GENSYS [8], NUCARS [9,10], SIMPACK [11], CONTACT [12], and VAMPIRE [13]. Such best-known software packages effectively help product manufacturers in efficiently applying sophisticated engineering methods with the needed simulation and services. In general, all systems were in close agreement on results of simulation but NUCARS and VAMPIRE had the fastest run times [14]. Software tools have a subsystem function for simulating interaction wheel and suspension with rail track by numerical integration of the equations of motion for dynamic analysis [15]. To achieve higher performance, it is advisable to use multi-core processors. Parallel processing is supported by ADAMS.

Modern multi-core architecture provides the developers with the different means of parallel software on hardware platforms with a symmetric multiprocessing architecture (SMP), in which all processors have direct and equal access to any cell in the shared memory [16]. Multithread programming tools are presented in the popular operating systems. However, for complicated dynamic systems with their associated issues, the development and effective implementation of multi-threaded models for multi-core systems are challenging.

The aim of the paper is to develop a parallel programming model to analyse the influence of parameters on the dynamic characteristics of vehicle control systems in multicore computer. Such a parallel solver for numerical integration of the equations of motion must be shared among cores of computer.

The rest of this paper is organized as follows: Firstly, Section 2 introduces a model of vehicle rail dynamic system. Secondly, Section 3 presents MATLAB simulation of the dynamic system model. Then, Section 4 describes the program model. Finally, Section 5 concludes the presented research and suggests possible extension for future work.

## II. MODEL OF VEHICLE RAIL DYNAMIC SYSTEM

Operation of the high-speed railways has high requirements for the dynamic qualities of locomotives and railway rolling stock [3]. The subsystem of numerical integration of the equations of dynamics is an important part of the system design and study models of railway transport and is based on the mechanics of wheel-rail interaction as described by the equations of motion [17, 18].

An essential part of components vehicle dynamics is the mass-spring-damper system with two-degree-of-freedom. Figure 1 shows a scheme of a vehicle traveling on a rough road. Wagon and bogies have traditionally been replaced by the lumped masses in the simulation of dynamic characteristics of the train, as shown in Figure 1-b. The suspension bogie is achieved using two springs and a damper connected in parallel as shown in Figure 1-a. The degree of freedom of the system is determined by the number of independent generalized coordinates $z_k$, $z_1$ and $z_2$ associated with vertical movements of the rail bogies. Analysis of the vertical oscillation presented during the movement includes the wagon rotation around the longitudinal axis $\varphi_k$, with corresponding moment of inertia $J_k$.

The mathematical model of the system is given by an ordinary differential equation. Using the principle of D'Alembert-Lagrange, the system of differential equations can be written in the following matrix form [19]:

$$Iz'' + Dz' + Sz = F \qquad (1)$$

where I is the inertia matrix, D is the dissipative matrix, S is the stiffness matrix, F is a vector of generalized forces depending on the speed and the geometric irregularities (the roughness matrix), $z''$ is the accelerations column vector, $z'$ is the velocity vector, and $z$ is the displacement vector considering the generalized coordinates.

Matrix notation of equations is useful in dynamic models with a large number of degrees of freedom. The matrix form of the model can be reduced to a flat form [20-22]. In this case, the wagon, oscillations of the first bogie and oscillations of the second bogie equations are as follows:

The equations of the wagon:

$$m_k * z_k'' + b_k * (2 * z_k' - z_1' - z_2') + c_k * 2 * (z_k - z_1 - z_2) = 0 \qquad (2)$$

$$J_k * \varphi_k'' + a_k * b_k * (2 * a_k * \varphi_k' - z_1' + z_2') + a_k * c_k * (2 * a_k * \varphi_k - z_1 + z_2) = 0 \qquad (3)$$

The equations oscillations of the first bogie:

$$c_b * z_1'' - b_k * (z_k' - z_1' + a_k * \varphi_k') - c_b * (z_k - z_2 + a_k * \varphi_k) + 2 * b_b * z_1' + 2 * c_b * z_1 =$$
$$b_b * (\eta_1' + \eta_2') + c_b * (\eta_1 + \eta_2) \qquad (4)$$

The equations oscillations of the second bogie:

$$m_b * z_2'' - b_b * (z_k' - z_2' + a_k * \varphi_k') -$$
$$c_b * (z_k - z_2 + a_k * \varphi_k) + 2 * b_b * z_2' + 2 * c_b * z_2 =$$
$$b_b * (\eta_3' + \eta_4') + c_b * (\eta_3 + \eta_4) \qquad (5)$$

where $m_k = 57$ т is the mass of the wagon, $J_k = 70$ is the moment of inertia of the wagon, $m_b = 9$ т is the mass of the bogie, $a_k = 3{,}725$ m denotes half base of wagon, $a_b = 1{,}5$ meters denotes half base of bogie, $c_b = 3040$ kN/m is the stiffness of the bogie, $b_b = 30$ kNsec/m is the damping of the bogie, $c_k = 2660$ kN/m is the stiffness of the wagon, $b_k = 100$ kNsec/m is the damping of the wagon, $z_i$, $z_i'$, $z_i''$, $\varphi_k$, $\varphi_k'$, $\varphi_k''$ are the generalized coordinates displacement and rotation as well as their time derivatives, and ηj(t) is the disturbance from the surface of the rail on pair of j-wheels. Simulation of tracks may include various track geometries. Assume that the surface of the road can be approximated as a sine wave [20]:

$$\eta(t) = a_1 * \sin(w * t) + a_2 * \sin(3 * w * t), \qquad (6)$$

where $w = 2\pi V / L = 5{,}027$ is the disturbance frequency, V = 20m/sec is the velocity, $L = 25$m is the path length.

External disturbances to the system with transport delay $\eta_i = \eta(t-\tau_i)$, $\tau = \frac{2a}{V}$ are determined by the function with geometrical dimensions and the movement velocity. The displacement $\eta_i$ and acceleration $\eta_i'$ are determined with the delay at the moments of time $2a_1/V$, $2a_2/V$, $2(a_1+a_2)/V$ for the corresponding wheels, where $a_1 = 0.005$m and $a_2 = 0.002$m are the amplitudes of flatness.

### III. MATLAB SIMULATION OF THE DYNAMIC SYSTEM MODEL

MATLAB contains solvers for stiff and non-stiff problems [23]. To verify the digital multi-threaded multi-core model below, a model for the system of differential equations (1)-(6) was formulated for MATLAB simulating the dynamic response of a railroad vehicle based on specified track conditions.

A complete description of the process of constructing a mathematical model involves the following blocks: initial data, the initial conditions for differential equations, integration of differential equations, disturbance from the rail, nonlinear functions, conversion of the output data and plotting the results. For integrating the system equations, an adaptive method implemented using the MATLAB function ode45 is employed. The differential equations of the dynamical system (1)-(6) are converted to a system of first order equations by reducing the order of the derivative before the integration by Runge-Kutta method. The equations of the dynamical system are then converted to a form suitable for integrating with MATLAB as follows:

$$D(1) = x(2)$$
$$D(2) = \frac{1}{m_1} * \Big(b_1\big(n_{11} + n_{22} - 2*x(2)\big) + c_1 * \big(n_1 + n_2 - 2*x(1)\big) + b_2 * \big(x(6) - x(2) + a_2*x(8)\big) + c_2 * \big(x(5) - x(1) + a_2*x(7)\big)\Big)$$

$$D(3) = x(4)$$
$$D(4) = \frac{1}{m_1} * \Big(b_1 * \big(n_{33} + n_{44} - 2*x(4)\big) + c_1 * \big(n_3 + n_4 - 2*x(3)\big) + b_2 * \big(x(6) - x(4) - a_2*x(8)\big) + c_2 * \big(x(5) - x(3) - a_2*x(7)\big)\Big)$$

$$D(5) = x(6)$$
$$D(6) = \frac{1}{m_2} * \Big(b_2 * \big(x(2) + x(4) - 2*x(6)\big) + c_2 * \big(x(1) + x(3) - 2*x(5)\big)\Big)$$

$$D(7) = x(8)$$

$$D(8) = \frac{a_2}{j_2} * (b_2 * (x(2) - x(4) - 2*x(8)*a_2) + c_2(x(1) - x(3) - 2*x(7)*a_2)) \quad (7)$$

The integration block of the differential equations has $m_1 = m_b$, $m_2 = m_k$, $b_1 = b_k$, $c_2 = c_k$, $a_2 = a_k$ and $b_2 = b_b$.

Figures 2-6 illustrate the simulation results of system in MATLAB environment.

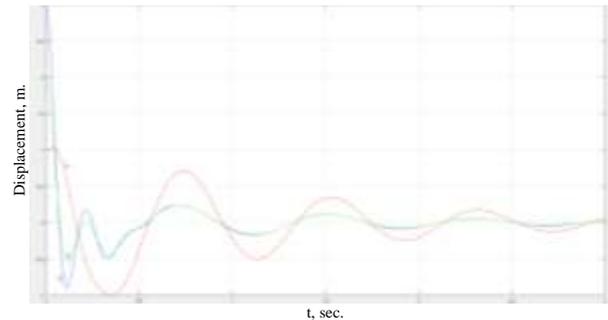

Figure 2. Graph of displacements $z_2$, $z_4$, $z_6$ in time in a multithreaded model (carcass: red, 1st bogie: blue and 2nd bogie: green).

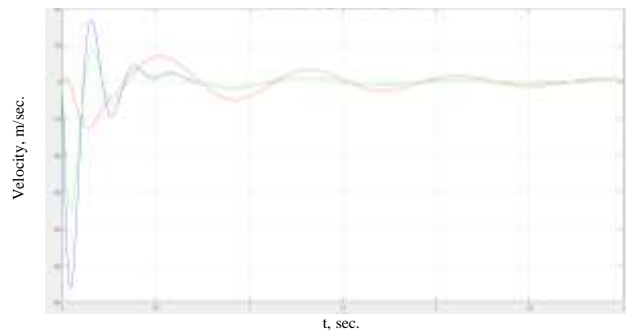

Figure 3. Graph of velocity $z'_2$, $z'_4$, $z'_6$ in time in a multithreaded model (carcass: red, 1st bogie: blue and 2nd bogie: green).

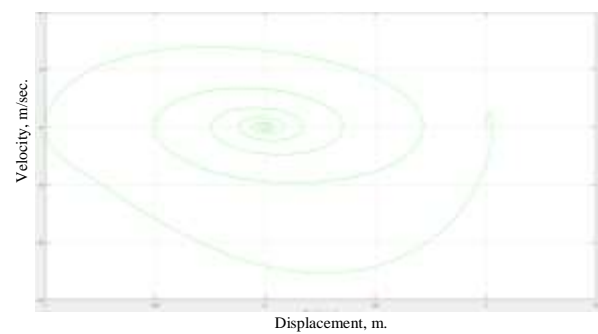

Figure 4. Carcass Phase diagram.

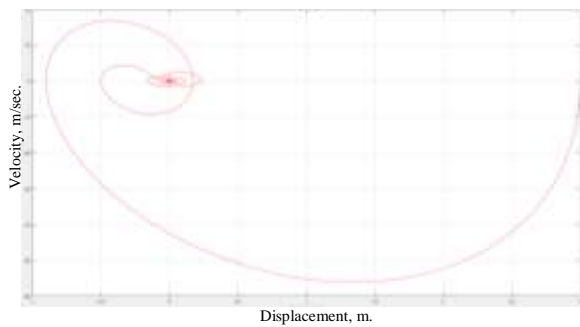

Figure 5. 1st bogie phase diagrams

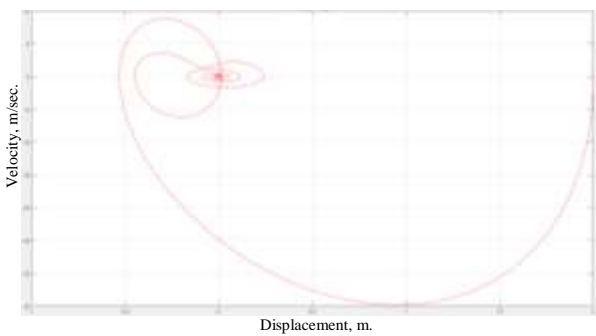

Figure 6. 2nd bogie phase diagrams.

Based on the dynamic analysis for studying the dynamic response of the bogies sections, the model exhibits a stable response during the motion.

## IV. PROGRAM MODEL

For efficient implementation, parallel models need to take into account the limitations of data and information threads. The simultaneous control of actions leads to problems of four types: synchronization, interaction, load balancing and scalability [24].

In the first stage of program model development, a distribution method of equations is chosen which is similar to the schema of analog computer. Each equation model is distributed on a separate core. The computational scheme corresponds to the system of equations (1-6) is reduced to solution of 1st order differential equations. The transformed system is numerical solved using the Runge-Kutta method.

The Windows operating system provides computer system operation with a symmetric multiprocessing architecture (SMP). In this case, all processors have direct and equitable access to the cells of the shared memory. Intel Inc. ® offers different software resources helping to effectively use the optimum thread technology [25]. The basis for the creation of a dynamic model of the system is thread-level parallelism. When implementing the model, it is possible to use low-level interface functions such as: API for Microsoft Win32, API POSIX-thread and API *OpenMP*. In the process, 4 threads (numbered as 1, 2, 3 and 4 in Figure 7) are created using *CreateThread* Microsoft API functions for numerically solving the involved system of equations (1 & 5, 2 & 6, 3 and 4, respectively).

In Figure 7, for each of the four processor cores and threads, an affinity mask is created using *SetThreadAffinityMask* function. Threads are created with the CREATE_SUSPENDED parameter and started with calling the *ResumeThread* function. To minimize the effect of the overhead expenses of the operating system, the *SetThreadPriority* function is used with a priority value THREAD_PRIORITY_ABOVE_NORMAL.

Finishing simulation is provided by the *WaitForMultipleObjects* synchronization function at the end of numerically solving all threads in the respective cores. Performing thread ends when the thread returns the control to the process. The *CreateThread* function creates threads in the context of the modeling process, from which they are called. Since the model threads are located within the same process, they share the address space (one memory context) and have access to the global variables [26,27]. The system provides different physical addresses for each thread in the range of the virtual addresses section for local variables of threads. This distribution model of equations among the cores has led to the fact that the variables to be exchanged among the threads are global variables of the application.

The main problem of multi-threaded programming is the implementation of critical sections so that the threads are not simultaneously used. When implementing the synchronization mechanisms a recommendation is adopted to minimize the number of read-write locks of variables. This mechanism provides simultaneous reading from multiple threads, but the saving is allowed only for one thread [28].

The size of the critical section becomes less important. Also the complexity of the synchronization data are transferred to the informational thread through the interaction caches L1, L2, L3 as shown in Figure 7 [29]. Modern Intel processors have a three-level cache memory subsystem [30]. Caches of different levels in the processor perform a variety of tasks. The 1st level L1 cache belongs only to a particular processor core and consists of a data cache (L1D) and cache (L1I). The 2nd level L2 cache is available for pairwise cores. The 3rd level L3 cache is shared among all processor cores. The L3 cache is inclusive relative to L1 and L2 caches, that is, always duplicated contents of L1 and L2 caches. Access to the cached data does not occupy memory bus bandwidth, but creates a coherence problem [31]. As a hardware platform for a dynamic system is used, a 4-core Intel processor includes a 32 KB L1 cache, 256 KB L2 cache and 8 MB L3 cache. The problem of cache coherence occurs at exchange of the variables $z_k$, $z_1$, $z_2$, $\varphi_k$ between threads. With alignment method (padding), variables data sharing between the cores were taken into consideration by the system time measurement for each core by the API-functions *QueryPerformanceFrequency* and *QueryPerformanceCounter*.

Multithreaded code performance requires validation [28]. Concurrency visualizer extension for Visual Studio 2015 is used for this purpose. The tool provides graphical, tabular, and textual data that show the temporal relationships between the threads in the program and the system as a whole. Concurrency visualizer can be used to locate performance bottlenecks, CPU underutilization, thread contention, cross-core thread migration, synchronization delays, areas of overlapped I/O, and other information.

The performance of the program model has been examined based on how the thread execution was mapped to the logical processor cores. As already mentioned, each thread has been executed in the separate core following the adopted approach. Figure 8 proves that this task has been successfully achieved. The graph shows logical cores on the y-axis and time on the x-axis. All threads in the graph have unique colors. There are two stages of the program. The 1st stage is a preparation stage. At the preparation stage, 4 threads are created and configured. The number of cross-core context switches for 3 threads is 1 and for one thread is 0. It is because all threads are created in a core, then they may move to another core and after that they are executed only on the specified core. The 2nd stage is an execution stage. For execution, the problem of cross-core migration is completely eliminated due to the correct use of thread affinity. Also it indirectly guarantees optimized cache performance.

In order to fully assess the program performance, the reasons of context-switches also should be explained. The most common reasons of context-switches are: blocking synchronization primitive, expiring of a threads' quantum, and blocking of I/O request. Threads tab of the concurrency visualizer shows the information given in Figure 9. 0% of I/O indicates that we fully excluded I/O blocking from the program. 1% of preemption means that the operating system affects the execution of the program, but the effect value is too small to make significant changes. The distribution of work across all parallel threads is equable. These results guarantee high performance of the application. Synchronization time is 56% that is usually a poor indicator. But in this case, the execution behavior proves that there are no locks in the program. The synchronization blocking can be explained by a different complexity of the equations that are solved in different threads. Each thread on every step will wait for the slowest thread to complete. That is why the overall working time for all threads is the same. The difference in execution time among threads is not significant due to the "Per Thread Summary" diagram of Figure 9. The synchronization problem is a disadvantage of the chosen model implementation approach. Despite this, all obtained profiling results indicate that the proposed application has high performance as mentioned above.

The results of numerical experiments on a multi-core multi-threaded model for speeds of 20-150 km/h have coincided with the results of MATLAB simulations. Analysis of the profiling of the dynamic model leads to the following recommendations for the implementation of the exchange with the memory by a cache in the model:

- Avoid conflict of cache lines.
- Perform caching optimization at the level of data structures and algorithms for processing.
- Consider a natural alignment of the stack variables (local), regardless of their size, but most compilers are aligned on addresses that are multiples of four.

V. CONCLUSION AND FUTURE WORK

A model has been developed that can be considered as an architectural solution for system organization designed for analysis of transient problems with different operating modes. Specifically, it is also effectively suitable for the design of the rolling stock. A model for the given system of differential equations was prepared for MATLAB implementation and used for validation of the HIL model. Moreover, hardware features recommendations for multi-core systems implementations have been developed. In addition, the possibilities of Windows operating system-based implementations on SMP systems have been studied with subject to the restrictions on the synchronization and communication between threads. Thread-level parallelism has been employed for creating a dynamic model of the system. For processors that support low-level multithreading technology, Microsoft Windows family of systems can be used for creating HIL models. Furthermore, the paper discusses the need to analyze the quality of parallel software implementation models by Microsoft and Intel development tools. It has been shown that the use of the Concurrency Visualizer tool for search on "bottlenecks" models (locking, data races, etc.) can improve the performance of parallel model. Generally, the proposed development method and model can be used for design and computational projects and also provide a tool for educational purposes and further application.

A suggested future scope of this research work is to seek about a high efficiency parallel solver for numerical integration of the equations of motion of rolling stock with maximized performance of the parallel solver. The extended research should develop effective schemes for distributing the model equations among multi-processor cores and analysis of a wide range of development tools and platforms (also real-time) for solver implementation applied to efficient practical applications.

ACKNOWLEDGMENT

The author would like to thank Taibah University for supporting this research.REFERENCE

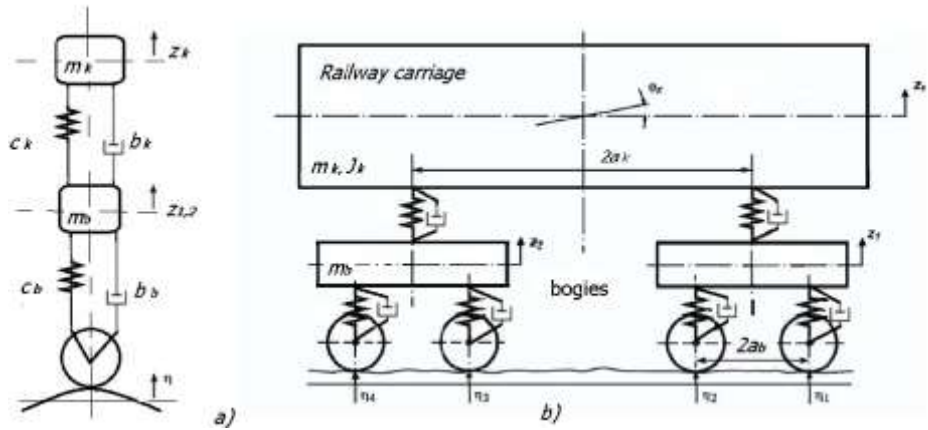

Figure 1. Schema of vertical oscillation of bogies: a) For a single wheel. b) For two bogies.

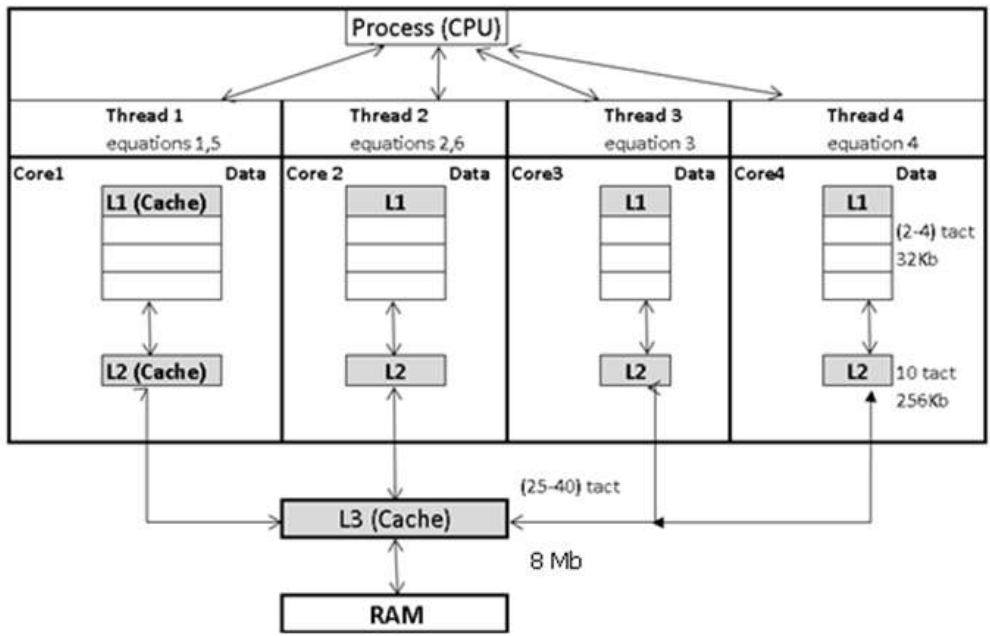

Figure 7. Platform structure hardware for a dynamic system.

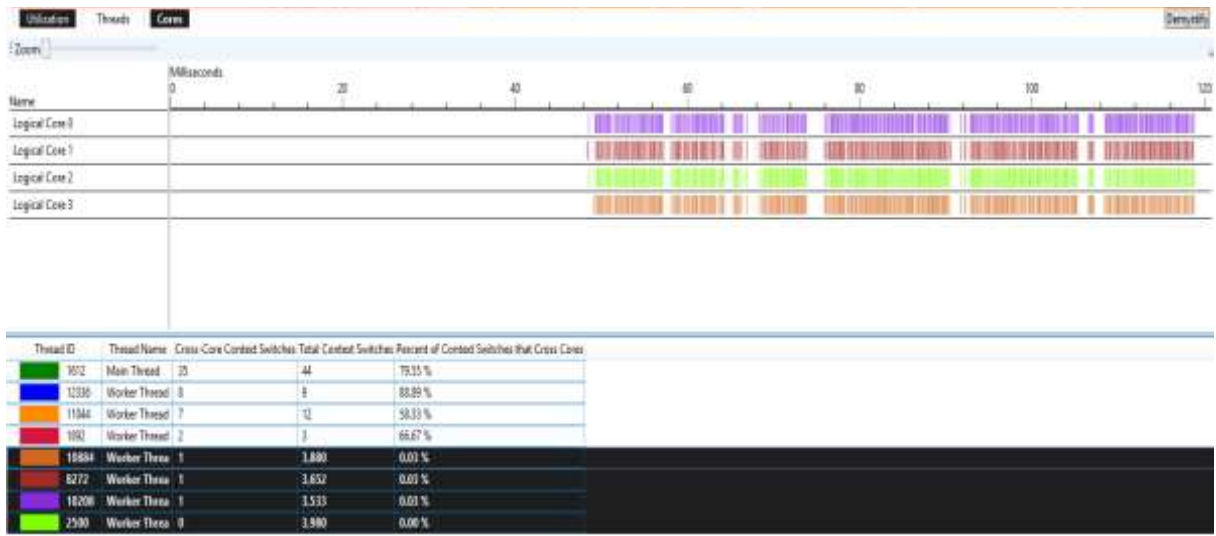

Figure 8. Mapping of model threads to the logical processor cores.

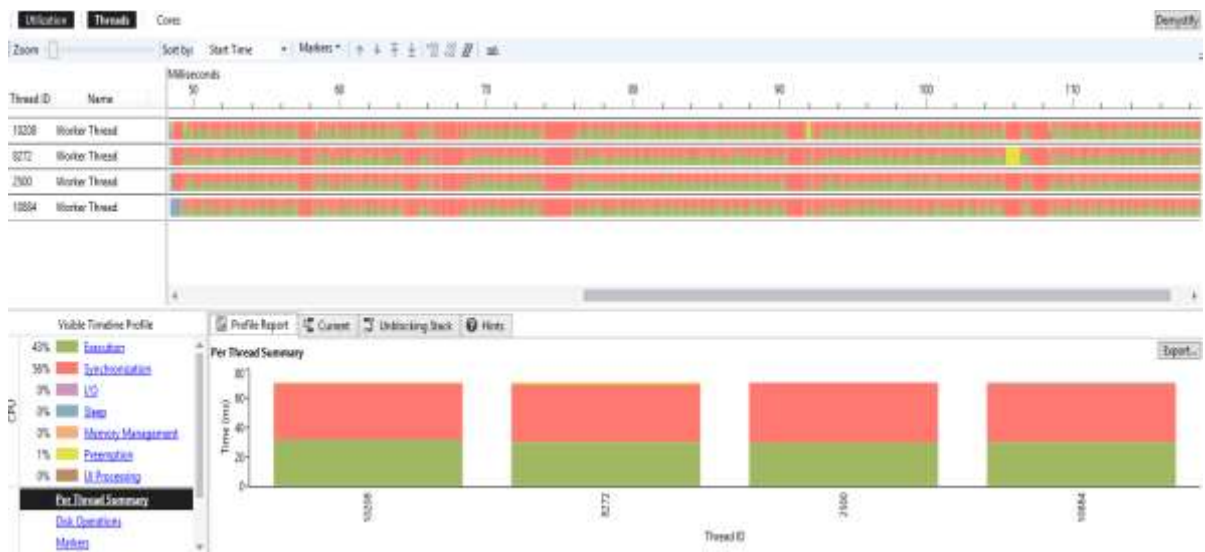

Figure 9. Behavior of program threads execution.